\documentclass[aps,twocolumn,showpacs]{revtex4}
\usepackage{graphicx}
\include{amssym}
\def\PSfig#1#2{\centerline{\scalebox{#1}{\includegraphics{#2}}}}

\begin{document}

\title{Lowering the critical temperature with eight-quark interactions} 

\author{A. A. Osipov}
\affiliation{Dzhelepov Laboratory of Nuclear Problems, 
         Joint Institute for Nuclear Research, 
         141980 Dubna, Moscow Region, Russia}
\author{ B. Hiller, J. Moreira, A. H. Blin, J. da Provid\^encia}
\affiliation{Centro de F\'{\i}sica Te\'{o}rica, Departamento de
         F\'{\i}sica da Universidade de Coimbra, 3004-516 Coimbra, 
        Portugal}

\begin{abstract}
It is shown that eight-quark interactions, which are needed to
stabilize the ground state of the combined three flavor Nambu -- 
Jona-Lasinio and 't Hooft Lagrangians, play also an important role
in determining the critical temperature at which transitions occur
from the dynamically broken chiral phase to the symmetric phase. 
\end{abstract}

\pacs{11.10.Wx, 11.30.Rd, 11.30.Qc}
\maketitle

For a long time already, a phase transition which almost certainly
occurs in hadronic matter at finite temperature is at the centre of 
attention of many researches and reviews 
\cite{Wilczek:1984,Brown:1990,Ortmanns:1996,Lenaghan:2000}. It 
may be very important in the context of heavy-ion collisions and the 
evolution of the early Universe. At present, experimental facilities 
with ultra-relativistic heavy-ion reactions aiming at studying the QCD 
phase diagram have triggered even more intense studies from the 
theoretical side, ranging from ef\mbox{}fective models to calculations 
on the lattice. Recent lattice results predict for vanishing chemical 
potential and massive quarks a non singular crossover for the QCD 
transition which lead to dif\mbox{}ferent critical temperatures, 
depending on the considered observables. The transition related to the 
renormalized chiral susceptibility yields $T_c=151(3)(3)$ MeV, whereas 
for instance the inclusion of the Polyakov loop increases this value
by $\sim 25$ MeV \cite{Aoki:2006}. 

An ef\mbox{}fective chiral model which has been widely used to extract
the low energy characteristics of hadrons involving the $u,d,s$ quarks 
is the well known Nambu -- Jona-Lasinio model \cite{Nambu:1961},
generalized to the $SU(3)_L\times SU(3)_R$ chiral symmetry and to
include the $U(1)_A$ symmetry breaking of low energy QCD in form of
the $2N_f$ determinantal interaction of 't Hooft ($N_f$ denoting the 
number of flavors) \cite{Hooft:1976}. This combined Lagrangian (we
denote it by NJLH) has been first analyzed in \cite{Bernard:1988} and 
\cite{Reinhardt:1988} to study dynamical breakdown of chiral symmetry
and related meson spectra in the vacuum, and since then in further
numerous applications 
\cite{Weise:1990,Takizawa:1990,Klevansky:1992,Hatsuda:1994,Bernard:1993}. 
In more recent calculations the model is used to obtain the
thermodynamic properties of hadrons and restoration of chiral and
$U(1)_A$ symmetries, both at zero and finite chemical potential 
\cite{Ruivo:2005}, and extended to include the colored diquark
channels, of importance at large baryonic densities \cite{Ruster:2005}.  
Calculations performed with the NJLH at zero chemical potential yield 
a critical temperature for the light quarks around $T_c\sim 200$ MeV,
where the condensate undergoes rapid changes. This value clearly 
overestimates the lattice prediction.

In the present letter, we argue that by extending the NJLH model to 
include interactions of eight quarks, the critical temperature is 
lowered considerably. The eight quark vertices have been introduced 
first in \cite{Osipov:2005b} to stabilize the scalar ef\mbox{}fective 
potential derived in the presence of $(2N_f=6)$ 't Hooft interactions 
\cite{Osipov:2005a} and have been discussed at length in 
\cite{Osipov:2006a}. 

We consider the following Lagrangian density
\begin{equation}
\label{efflag}
  {\cal L}_{eff} = \bar{q}(i\gamma^\mu\partial_\mu - m)q
          +{\cal L}_{4q} + {\cal L}_{6q}
          +{\cal L}_{8q}+\ldots\, .
\end{equation}
Quarks $q$ have color $(N_c=3)$ and f\mbox{}lavor $(N_f=3)$ indices 
which are suppressed here. We suppose that multi-quark interactions 
${\cal L}_{4q}$, ${\cal L}_{6q}$, ${\cal L}_{8q}$ are 
\begin{eqnarray}
\label{L4q}
  {\cal L}_{4q} &\! =\! &\frac{G}{2}\left[(\bar{q}\lambda_aq)^2+
                    (\bar{q}i\gamma_5\lambda_aq)^2\right], \\
\label{Ldet}
  {\cal L}_{6q} &\! =\! &\kappa (\mbox{det}\ \bar{q}P_Lq
                            +\mbox{det}\ \bar{q}P_Rq), \\
  {\cal L}_{8q} &\! =\! &{\cal L}_{8q}^{(1)} + {\cal L}_{8q}^{(2)}, 
\end{eqnarray}
where the f\mbox{}lavor space matrices $\lambda_a,\ a=0,1,\ldots ,8,$ 
are normalized such that $\mbox{tr} (\lambda_a \lambda_b ) = 2
\delta_{ab}$, and $\lambda_0=\sqrt{\frac{2}{3}}\, 1$, $\lambda_a$ at 
$a\neq 0$ are the standard $SU(3)$ Gell-Mann matrices. Chiral
projectors are $P_{L,R}=(1\mp\gamma_5)/2$, the determinant in the 't 
Hooft Lagrangian is over f\mbox{}lavor indices. The eight-quark 
interactions are given by 
\begin{eqnarray}    
   {\cal L}_{8q}^{(1)}&\!\! =\! & 
   8g_1\left[ (\bar q_iP_Rq_m)(\bar q_mP_Lq_i) \right]^2, \\ 
   {\cal L}_{8q}^{(2)}&\!\! =\! & 
   16 g_2(\bar q_iP_Rq_m)(\bar q_mP_Lq_j) 
   (\bar q_jP_Rq_k)(\bar q_kP_Lq_i). 
\end{eqnarray}
Here the summation over repeated f\mbox{}lavor indices is assumed. One 
sees that the most general spin zero eight-quark interaction is 
composed of two chiral invariant combinations of products of four
quark bilinears with couplings $g_1,g_2$, of which one, 
${\cal L}_{8q}^{(1)}$, is OZI violating. 

The model has eight parameters: the couplings of multi-quark
interactions $G, \kappa , g_1, g_2$; three masses of current quarks,
$m_i$, and the cutof\mbox{}f $\Lambda$ (the model being not 
renormalizable).

Since there is some freedom in the model parameter choices that in the 
realistic case with $SU(2)_I\times U(1)_Y$ f\mbox{}lavor symmetry and 
non zero current quark masses, $m_u=m_d\ne m_s$, lead to qualitatively 
similar spectra for the low lying pseudoscalars and scalars, we chose 
as input several sets of parameters at $T=0$. 

We put furthermore all current quark masses to zero to have clean 
signatures of the phase transitions. This will not af\mbox{}fect the 
trends we are about to discuss in connection with the pivot 
temperature values $T_i$ (see below) obtained with and without the 
stabilizing eight-quark interactions. On the other hand, this 
simplif\mbox{}ies essentially our analysis. 

Let us recall that at $T=0$ the ef\mbox{}fective potential of the 
model as a function of the constituent quark mass, $M$, in the 
$SU(3)$ limit is 
\cite{Osipov:2006a}
\begin{eqnarray}
\label{Vh}
   V(M)&\! =\!& \frac{h^2}{16}\left( 12G+\kappa h 
           +\frac{27}{2}\rho h^2\right) \nonumber\\
   &\! -\!&\frac{3N_c}{16\pi^2}\left[ M^2J_0(M^2)
       +\Lambda^4 \ln\left(1+\frac{M^2}{\Lambda^2}\right)\right]\! ,
\end{eqnarray}
with $\Lambda$ being an ultraviolet covariant cutof\mbox{}f in the quark 
one-loop diagrams, $\rho \equiv g_1+\frac{2}{3}g_2$, and 
\begin{equation}
\label{j0}
   J_0(M^2)=\Lambda^2- M^2\ln\left(1+\frac{\Lambda^2}{M^2}\right).
\end{equation}

The function $h(M)$ is a real solution of the stationary phase
equation related to the integration over bosonic auxiliary variables 
(for details see {\it e.g.} \cite{Osipov:2005b})
\begin{equation}
\label{spa1}
   M+Gh+\frac{\kappa}{16}h^2+\frac{3}{4}\rho h^3=0. 
\end{equation}
This equation yields a one-to-one real-valued mapping $M\to h$ in the 
parameter region
\begin{equation}
\label{stable}
     \rho >0, \hspace{0.4cm} 
     G>\frac{1}{\rho}\left(\frac{\kappa}{24}\right)^2
\end{equation}
which defines the stability conditions in the $SU(3)$ limit of the 
ef\mbox{}fective potential. 

The gap equation, obtained from the extremum condition $dV/dM=0$, 
relates further the function $h(M)$ to the quark condensate as
\begin{equation}
\label{cond}
   h(M)=-\frac{N_c}{2 \pi^2} M J_0(M^2)
                  =2\big <0|\bar qq|0\big >.
\end{equation}

The curvature of the ef\mbox{}fective potential $V(M)$ at the 
origin, $M=0$, in the chiral limit is determined at $T=0$ by the 
value of 
\begin{equation}
\label{tau}
   \tau=\frac{N_c}{2\pi^2}G \Lambda^2,
\end{equation}
with $\tau>1$, $\tau=1$ and $\tau<1$ indicating a maximum, a saddle
point and a minimum, respectively. 

At finite temperature, after introducing the standard non-covariant 
three-momentum cutof\mbox{}f $\lambda$ for the quark loops, see {\it 
e.g.} \cite{Florkowski:1997}, one obtains for the ef\mbox{}fective 
potential
\begin{eqnarray}
\label{Vht}
  && V_T(M)=\frac{h^2}{16}\left( 12G+\kappa h 
           +\frac{27}{2}\rho h^2\right) \\
  &&-\frac{3N_c}{\pi^2}\int\limits_0^{\lambda} 
  dp p^2 \left[ 2T \ln\left(1+\exp\left( 
  \frac{E(p)}{T}\right)\right)-E(p)
  \right], \nonumber
\end{eqnarray}
with $E(p)=\sqrt{p^2+M^2}$ and $p$ denoting the magnitude of the 
3-momentum. 

The curvature at the origin is now a temperature dependent function
\begin{equation}
\label{taut}
   \tau (T)=\frac{2N_c}{\pi^2}G \lambda^2 F(t), \quad 
   t=\frac{T}{\lambda}\ , 
\end{equation}
\begin{equation}
\label{Ft}
   F(t)=-\frac{1}{2}+\frac{t^2}{6\pi^2} +2t 
   \ln\left(1+e^t\right) + 2t^2 L_2[-e^t],
\end{equation}
where 
\begin{equation}
\label{polylog}
   L_n[z]=\sum_{k=1}^{\infty} \frac{z^k}{k^n}
\end{equation}
denotes the polylogarithm function. 

\begin{figure}[t]
\PSfig{0.8}{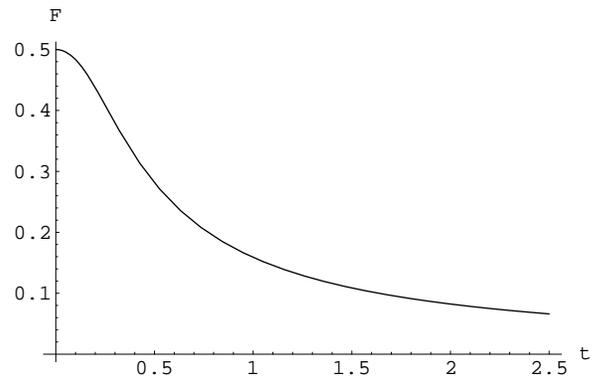}
\caption{The function F(t) related with the curvature of the effective 
         potential at the origin, in dimensionless units 
         $t=T/ \lambda$.} 
\label{myFigure1}
\end{figure}

\begin{figure}[t]
\PSfig{0.8}{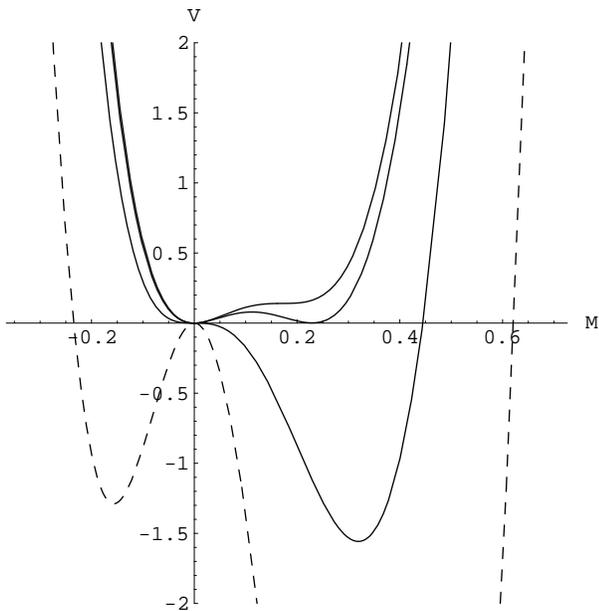}
\caption{The ef\mbox{}fective potential $V$ (in units (10 GeV)$^{-4}$) 
  for the parameter set (e) as function of M (in GeV). Full curves:
  lowest for $T_a$, middle for $T_d$ and upper for $T_s$. Dashed line: 
  T=0 case, the physical minimum occurs outside the plotting range.} 
\label{myFigure2}
\end{figure}

The function $F(t)$ is depicted in fig. 1. To guide the discussion we 
show in fig. 2 the temperature dependence of the ef\mbox{}fective 
potential for the parameter set (e) of Table 1, for $T=0$ (dashed line) 
and three ``critical" values $T_a,T_d,T_s$, introduced as follows. 

In the zero temperature limit the curvature reduces to $\tau (0)= 
N_c G \lambda^2/\pi^2$. From this we see that if we wish to ajust our 
zero of temperature curvature so that $\tau (0) =\tau$ we must choose
$\Lambda^2=2\lambda^2$. The function $F(t)$ decreases monotonically to 
zero as $T\to\infty$, starting from $F(0)=1/2$ and therefore, if the 
curvature $\tau> 1$ at $T=0$, the system will undergo at some critical 
temperature $T=T_a$ a change of curvature at the origin, starting from 
a saddlepoint at $T_a$ and entering a region of local stability near
this point. Increasing further the temperature the system undergoes 
changes reminiscent of a first order transition, passing through a 
configuration with two degenerate minima at $T=T_d$ until at some 
higher temperature $T_s$ a further saddlepoint appears, and for 
$T>T_s$ only the symmetric phase survives. Obviously if $\tau <1$ the 
system is at $T=0$ already in the coexistence or even in the symmetric 
phase. 

In the following we select a few parameter sets that have been
obtained previously with and without inclusion of the eight-quark 
interactions to determine the mass spectra of low lying pseudoscalar 
and scalar nonets, related weak decay constants and quark condensates.
In Table 1, the model parameters $\Lambda, G, \kappa, g_1, g_2$, are 
taken from \cite{Osipov:2006b}, set (a), and from \cite{Osipov:2006a}, 
sets (d,e,f), at $T=0$. This input is kept fixed in the finite 
temperature evolution of the ef\mbox{}fective potential. One sees that 
the eight-quark interactions present in sets (e,f) reduce considerably 
the temperatures $T_a,T_d,T_s$, as compared to the cases with $g_1=g_2=0$.  
In the last set (f) the curvature $\tau$ is slightly below 1, so the 
system has already a minimum at the origin at $T=0$. 

\vspace{0.5cm}
\noindent
{\small TABLE I. 
Parameters of the model at T=0: $G$
(GeV$^{-2}$), $\Lambda$ (MeV), $\kappa$ (GeV$^{-5}$), $g_1,\, g_2$ 
(GeV$^{-8}$). Indicated are also the temperatures $T_a,T_d,T_s$ 
(MeV)(see text).}

\noindent
\begin{tabular}{cccccccccc}
\hline \hline
Sets &$\ \ \Lambda$ &\ \ G   &\ \ $-\kappa$ &\ \ $g_1$ &\ \ $g_2$ 
     &\ \ $T_a$     &\ \ $T_d$ &\ \  $T_s$
\\ \hline
a &\ \ 820 &\ \ 13.5 &\ \ 1300 &\ \ 0  &\ \ 0   &\ \ 192 &\ \ 202 
  &\ \ 204 \\ 
d &\ \ 839 &\ \ 12.16 &\ \ 1082  & \ \ 0    &\ \ 0   &\ \ 174 
  &\ \ 183 &\ \ 185 \\ 
e &\ \ 839 &\ \ 11.28 &\ \ 1083  &\ \ 1500  &\ \ 327 &\ \ 143 
  &\ \ 161 &\ \ 163 \\ 
f &\ \ 839 &\ \ 8.92  &\ \ 1083  &\ \ 6000  &\ \ 327 &\ \  -  
  &\ \ 111 &\ \ 135\\ 
\hline \hline
\end{tabular}   
\vspace{0.5cm}

Let us discuss now the reasons for the obtained decrease of $T_c$ . The 
underlying mechanism is quite simple. The critical temperature is
correlated with the value of $\tau (0)$: the closer $\tau (0)$ is to 1, the
lower $T_c$. The eight-quark interactions decrease the ef\mbox{}fective
value of $G \Lambda^2$ (to see this one should consider the mass spectrum of
meson states \cite{Osipov:2006a}), and therefore lower the
transition temperature. In the parameter sets shown we have: $\tau
(0)=1.4$ (a), $\tau (0)=1.3$ (d), $\tau (0)=1.2$ (e) and 
$\tau (0)=0.95$ (f). 

The lowering of the critical temperature $T_c$ has also been recently 
observed in the $N_f=2$ case with eight-quark interactions 
\cite{Kashiwa:2006}.         

We conclude that the presence of the eight-quark interactions in the 
hadronic vacuum results in an observable ef\mbox{}fect which deserves a
more detailed study. The arguments presented above lead to the 
expectation that the tendency will be the same when one considers the 
realistic case with physical current quark masses. Work in this
direction is in progress.

\vspace{0.5cm}
{\bf Acknowledgements}
This work has been supported in part by grants provided by 
Funda\c c\~ao para a Ci\^encia e a Tecnologia, POCI/FP/63412/2005, 
POCI/FP/63930/2005 and SFRH/BD/13528/2003. This research is part of 
the EU integrated infrastructure initiative Hadron Physics project 
under contract No.RII3-CT-2004-506078. 



\begin{thebibliography}{99}

\bibitem{Wilczek:1984} R. D. Pisarski, F. Wilczek, Phys. Rev. D {\bf
    29}, 338 (1984).
\bibitem{Brown:1990} F. R. Brown {\it et al.,} Phys. Rev. Lett. {\bf
    65}, 2491 (1990).  
\bibitem{Ortmanns:1996} H. Meyer-Ortmanns, Rev. Mod. Phys. {\bf 68}, 
    473 (1996).  
\bibitem{Lenaghan:2000} J. T. Lenaghan, D. H. Rischke,
    J. Schaffner-Bielich, Phys. Rev. D {\bf 62}, 085008 (2000), 
    nucl-th/0004006.
\bibitem{Aoki:2006} Y. Aoki, G. Endrodi, Z. Fodor, S.D. Katz,
    K. K. Szabo, Nature {\bf 443}, 675 (2006), hep-lat/0611014;
    {\it idem} Phys. Lett. B {\bf 643}, 46 (2006), hep-lat/0609068. 
\bibitem{Nambu:1961} Y. Nambu and G. Jona-Lasinio, Phys. Rev. {\bf 122}, 
    345 (1961); {\bf 124}, 246 (1961); V. G. Vaks and A. I. Larkin, 
    Zh. \'{E}ksp. Teor. Fiz. {\bf 40}, 282 (1961) [Sov. Phys. JETP
    {\bf 13}, 192 (1961)].
\bibitem{Hooft:1976} G. 't Hooft, Phys. Rev. D {\bf 14}, 3432 (1976);
    G. 't Hooft, Phys. Rev. D {\bf 18}, 2199 (1978).
\bibitem{Bernard:1988} V. Bernard, R. L. Jaf\mbox{}fe and
    U.-G. Meissner, Phys. Lett. B {\bf 198}, 92 (1987);   
    V. Bernard, R. L. Jaf\mbox{}fe and U.-G. Meissner, Nucl. Phys. B
    {\bf 308}, 753 (1988). 
\bibitem{Reinhardt:1988} H. Reinhardt and R. Alkofer, Phys. Lett. B 
    {\bf 207}, 482 (1988).
\bibitem{Weise:1990} S. Klimt, M. Lutz, U. Vogl and W. Weise,
    Nucl. Phys. A {\bf 516}, 429 (1990);
    U. Vogl, M. Lutz, S. Klimt and W. Weise, Nucl. Phys. A {\bf 516},
    469 (1990);
    U. Vogl and W. Weise, Progr. Part. Nucl. Phys. {\bf 27}, 195 
    (1991). 
\bibitem{Takizawa:1990} M. Takizawa, K. Tsushima, Y. Kohyama and K. 
    Kubodera, Nucl. Phys. A {\bf 507}, 611 (1990).
\bibitem{Klevansky:1992} S. P. Klevansky, Rev. Mod. Phys. {\bf 64},
    649 (1992).
\bibitem{Hatsuda:1994} T. Hatsuda and T. Kunihiro, Phys. Rep. {\bf 247},
    221 (1994).
\bibitem{Bernard:1993} V. Bernard, A. H. Blin, B. Hiller, U.-G. 
    Mei\ss ner and M. C. Ruivo, Phys. Lett. B {\bf 305}, 163 (1993), 
    hep-ph/9302245;
    V. Dmitrasinovic, Nucl. Phys. A {\bf 686}, 379 (2001), 
    hep-ph/0010047; 
    K. Naito, M. Oka, M. Takizawa and T. Umekawa, Progr. Theor. 
    Phys. {\bf 109}, 969 (2003), hep-ph/0305078.
\bibitem{Ruivo:2005} P. Costa, M.C. Ruivo, C.A. de Sousa, Yu. L. 
    Kalinovsky, Phys. Rev. D {\bf 70}, 116013 (2004), hep-ph/0408177; 
    {\it idem} Phys. Rev. D {\bf 71}, 116002 (2005), hep-ph/0503258. 
\bibitem{Ruster:2005} S. B. Ruster, V. Werth, M. Buballa, I. A. 
    Shovkovy, D. H. Rischke, Phys. Rev. D {\bf 72}, 034004 (2005),
    hep-ph/0503184. 
\bibitem{Osipov:2005b} A. A. Osipov, B. Hiller and J. da
    Provid\^encia, Phys. Lett. B {\bf 634}, 48 (2006), hep-ph/0508058.
\bibitem{Osipov:2005a} A. A. Osipov, B. Hiller, V. Bernard and A. H.
    Blin, Ann. of Phys. {\bf 321}, 2504 (2006), hep-ph/0507226; 
    B. Hiller, A.A. Osipov, V. Bernard, A.H. Blin, SIGMA 2:026 (2006),  
    hep-ph/0602165. 
\bibitem{Osipov:2006a} A. A. Osipov, B. Hiller, A. H. Blin and J. da
    Provid\^encia, Ann. of Phys. doi:10.1016/j.aop.2006.08.004 (2006),
    hep-ph/0607066.
\bibitem{Florkowski:1997} W. Florkowski, Acta Phys. Pol. B {\bf 28}, 
    2079 (1997).
\bibitem{Osipov:2006b} A. A. Osipov, B. Hiller, J. Moreira, A. H. Blin, 
    Eur. Phys. J. C {\bf 46}, 225 (2006), hep-ph/0601074. 
\bibitem{Kashiwa:2006} K. Kashiwa, H. Kouno, T. Sakaguchi, M. Matsuzaki, M. Yahiro, nucl-th/0608078. 

\end{thebibliography}
\end{document}